# Developing an edge computing platform for real-time descriptive analytics

Hung Cao, Monica Wachowicz, Sangwhan Cha
People in Motion Lab
University of New Brunswick
15 Dineen Drive, Fredericton, NB. E3B 5A3 Canada
{hcao3, monicaw, sangwhan.cha}@unb.ca

*Abstract*—The Internet of Mobile Things encompasses stream data being generated by sensors, network communications that pull and push these data streams, as well as running processing and analytics that can effectively leverage actionable information for transportation planning, management, and business advantage. Edge computing emerges as a new paradigm that decentralizes the communication, computation, control and storage resources from the cloud to the edge of the network. This paper proposes an edge computing platform where mobile edge nodes are physical devices deployed on a transit bus where descriptive analytics is used to uncover meaningful patterns from real-time transit data streams. An application experiment is used to evaluate the advantages and disadvantages of our proposed platform to support descriptive analytics at a mobile edge node and generate actionable information to transit managers.

## I. INTRODUCTION

The fast growing of data streams generated by the Internet of Mobile Things (IoMT) poses several challenges in pulling this data from IoMT devices to remote clouds. In particular, Lu et al. [1] points out that one critical challenge in building the next generation of intelligent transportation systems is related to the harsh communication environment inside and/or outside a moving vehicle. Solutions for vehicle-to-sensor, vehicle-to-vehicle, and vehicle-to-road infrastructure connectivity are stringently dependent on latency and reliability for controlling and monitoring purposes. Moreover, mobility applications usually require seamlessly computation, storage, and connection services over a vast geographical area (e.g. entire transit system), challenging the network communication technology used between sensors and the core network in terms of issues such as becoming unreliable and error-prone as well as requiring an extensive amount of data storage [2]. Finally, developing the appropriate analytical workflow for leveraging the data streams generated by moving vehicles in order to produce active information for decision making is still a challenge since the data streams might be unbounded, noisy, and incomplete.

Due to the unpredictable network latency, expensive bandwidth, resource-prohibitive and location-awareness concerns of the Internet of Mobile Things, edge computing emerges as a new paradigm that decentralizes the communication, computation, control and storage resources to the edge of the Internet [3]–[5]. For current transit network systems that produce a vast volume of data streams in real-time, edge computing brings the opportunity to analyze massive data streams related to any vehicle of a fleet at the time the data is being collected and deliver actionable information to support tactical and operational decisions of transit managers [6]. Some examples in public transit analytics include computing the actual maximum/minimum transit route length, service demand frequency, and predicting dynamic bus stops in real-time.

Despite the advantages of edge computing, no previous research work could be found in the literature on building edge analytics platforms for supporting transit network systems. This paper proposes an edge computing platform where mobile edge nodes are physical devices deployed on a transit bus where descriptive analytics is implemented to analyze real-time transit data streams.

The main contributions of this paper are summarized as follows.
- We present an edge computing architecture that supports mobile edge nodes, i.e. edge nodes that are deployed inside a vehicle, in our case, a bus belonging to a transit network system.
- We develop a set of descriptive analytics tasks to analyze real-time transit data streams at the edge of a network.
- We run a real-time experiment to evaluate the advantages and disadvantages of our proposed platform to support transit managers with actionable information.

The remainder of the paper is organized as follows. Section 2 describes the main paradigms of edge computing. Our edge computing platform to support descriptive analytics is described in Section 3. The results and discussions are presented in Section 4. Finally, the conclusions are drawn in Section 5.

## II. RELATED WORK

Currently, three main paradigms can be found in edge computing that can be described as Fog Computing [7], Mobile Edge Computing [8] and Mobile Cloud Computing [9]. *Fog Computing* was first introduced by Cisco as a bridge between IoMT devices and the cloud [7]. It supports a distributed computing model that provides services at highly geographically distributed fog nodes such as access points, switches, and routers. It is defined as *"a scenario where a huge number of heterogeneous (wireless and sometimes* *autonomous) ubiquitous and decentralized devices communicate and potentially cooperate among them and with the network to perform storage and processing tasks without the intervention of third-parties. These tasks can be for supporting basic network functions or new services and applications that run in a sandboxed environment. Users leasing part of their devices to host these services get incentives for doing so."* [10].

*Mobile Edge Computing* was introduced by Nokia Networks [11] with the aim of supporting a base station as an intelligent service hub that can collect real-time network data such as cell congestion and subscriber locations. The ETSI Industry Specification Group (ISG) has defined mobile edge computing as *"a concept that provides an IT service environment and cloud-computing capabilities at the edge of the mobile network, within the Radio Access Network (RAN) and in close proximity to mobile subscribers."* [8].

Similarly, *Mobile Cloud Computing* was proposed to overcome the shortage in computing power and storage capacity of mobile devices by leveraging the services of cloud computing to offload computation for these end devices [12], [13]. Khan et al. define mobile cloud computing as *"a service that allows resource-constrained mobile users to adaptively adjust processing and storage capabilities by transparently partitioning and offloading the computationally intensive and storage demanding jobs on traditional cloud resources by providing ubiquitous wireless access"* [14].

All these three paradigms aim to reduce the latency of sending the data from the IoMT devices to the core network, ensuring highly efficient network operation and service delivery as well as providing the edge analytics to offload the burden at the core network [9], [15]–[19]. But they have few different characteristics that play an important role in the selection of an appropriate computing platform for data analytics. Table 1 provides an overview of the main characteristics of these platforms in terms of bringing the computation ability for data analytics to the edge of a network.

Depending on the complexity of the network protocols and the high-performance computing needs, data analytics can be deployed at a fog node level whereas for some applications it might be more appropriate to deploy it centrally, typically hosted in a mobile cloud platform. Aazam and Huh [20] have already found that unnecessary communication not only burdens the core network, but also a data center in the cloud. Therefore, the research question still remains on how to determine what kind of application requires data analytics at edge, fog and/or cloud levels. Fig. 1 shows an attempt to mapping different applications to a local level (edge), aggregation level (fog), and cloud level. We expect a blend of these levels to support real-time data analytics, and the ability to manage stream data from local (i.e. edge) to cloud levels will be increasingly critical in the near future.

TABLE I. MAIN CHARACTERISTICS OF FOG COMPUTING, MOBILE EDGE COMPUTING, AND MOBILE CLOUD COMPUTING

|  | Fog computing | Mobile Edge Computing | Mobile Cloud Computing |
|---|---|---|---|
| *Owned & Managed by* | Any (Mobile Network Provider, Cloud Service Provider, Organizations, Individuals) | Mobile Network Provider | Private Organization, Individuals |
| *Target Users* | Any user | Available to mobile users | Specific users |
| *Network Access* | Any short and long range networks | Mobile networks | Any short range networks |
| *Geo-distribution* | Any location | Co-located with base station | Static location (data center, cloudlet) |
| *Mobility* | Yes | No | No |
| *Computing Environment* | Indoor / Outdoor | Indoor / Outdoor | Indoor |
| *Computing and Storage Capability* | Yes | Yes | Yes |
| *Latency (Delay)* | Low latency | Low latency | Ranges from low latency to high latency |
| *Edge Analytics* | Not deployed yet | Not deployed yet | Edge Analytics [21] |

One example includes the cloudlet concept previously proposed by Satyanarayanan and his colleagues [22]. Cloudlets are trusted, resource-rich, mostly stationary computers with well-connected to the Internet, offering bandwidth, computation, and storage resources to nearby mobile users. In [21], authors proposed GigaSight which is an Internet-scale repository of crowd-sourced video content. GigaSight's architecture is a federated system of VM-based cloudlets that perform video analytics at the edge of the network. Several scenarios have been envisaged to apply fog computing, including Augmented Reality (AR), Real-time video analytics, Mobile Big Data Analytics, Smart Grid, Smart Traffic Lights and Connected Vehicles, Decentralized Smart Building Control, Wireless Sensors and Actuators Networks [5], [19], [23]–[25]. Unfortunately, none of these scenarios has been actually implemented so far.

We are interested in exploring edge analytics to increase the efficiency of transit network systems of cities functioning in real-time. In particular, smart transit systems are generating a vast amount of sensing data that can generate contextual information needed in real-time for offering alternative modes of travel, reducing traffic congestion and improving the quality of life. To the best of our knowledge, there is no previous research work on deploying an edge analytical platform for a transit network system.

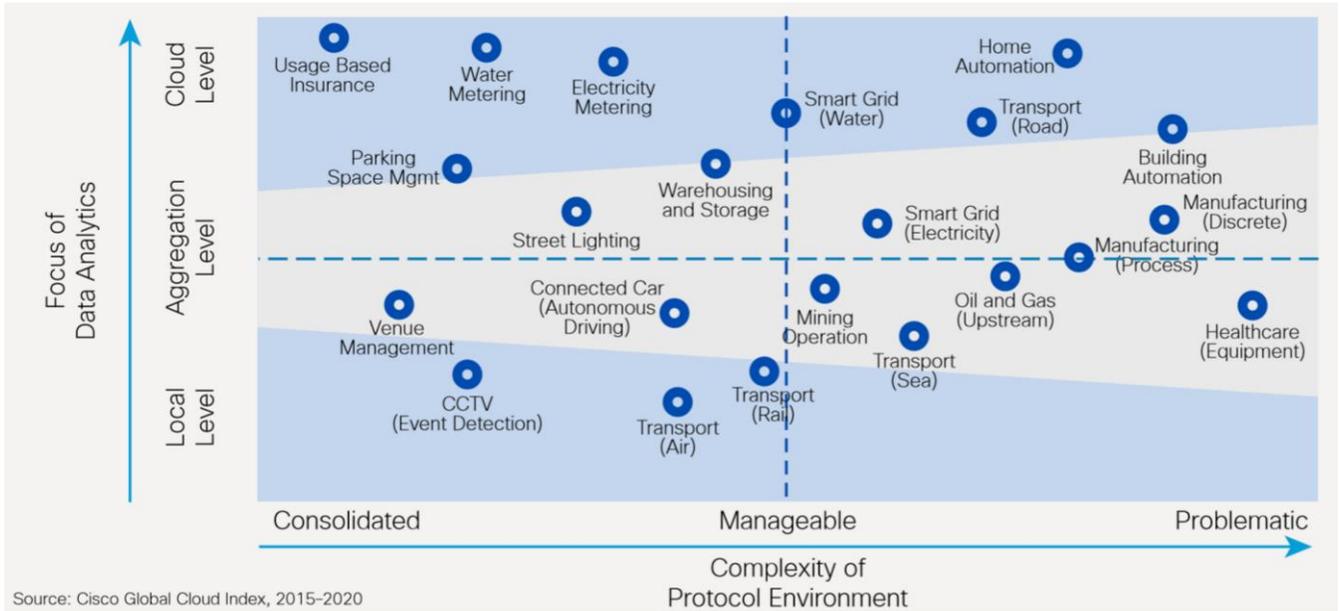

Figure 1. Application levels of complexity in edge analytics [27].

## III. EDGE ANALYTICS PLATFORM

Our edge computing platform consists of a mobile edge node. The main scenario can be described as a moving bus that generates real-time data streams (e.g. Automatic Vehicle Location (AVL) data feeds) which are fetched by a mobile edge node installed in this bus. The platform supports running different descriptive analytics tasks at the mobile edge node meanwhile the bus moves around a city. Once the analytical results are generated, they provide actionable information about what is happening to a moving bus. The stream data lifecycle in our platform supports data pre-processing tasks as well.

### A. Real-Time Data Streaming

The real-time data streams used in this research were provided by Codiac Transit which is responsible for delivering efficient transportation services with the aim to reduce private car dependency in the metropolitan area of Greater Moncton, Canada. Codiac Transit currently operates 30 regular routes from Monday to Saturday, some of which having additional evening and Sunday services. Each bus is equipped with GPS receivers which generate geographic coordinates of the location of a bus every 5 seconds for each bus trip. Moreover, telemetry data is generated with a total of 17 data fields, including route name, trip identifier, start and finish time of a trip. The location and telemetry data is transmitted to the mobile edge node using 4G LTE Dual SIM.

The bus route 51 was selected for evaluating our edge computing platform because it has the highest trip density during a day. For the purpose of explaining our real-time descriptive analytics we have used the data retrieved during the period from 02/14/2017 to 02/20/2017.

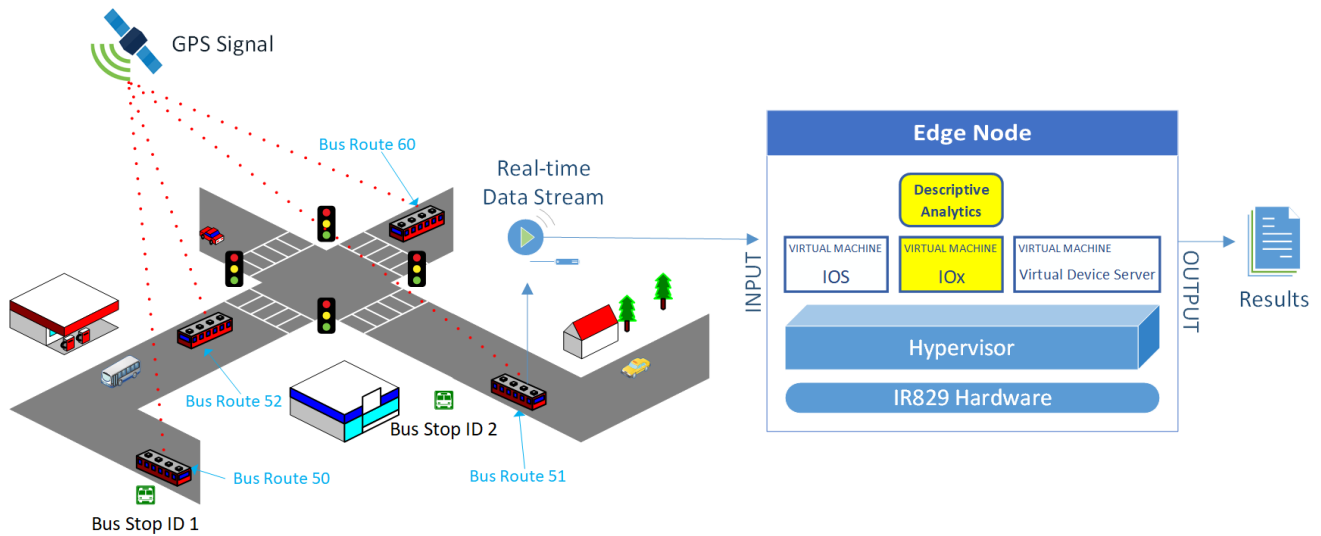

Figure 2. The overview of the edge computing architecture

### B. Mobile Edge Node

Every bus running on route 51 generates 17 data fields every 5 seconds which are sent to its mobile edge node installed inside the bus. In this experiment, the mobile edge node known as Cisco IR829 Industrial Integrated Services Router was used, having an Intel Atom Processor C2308 (1M Cache, 1.25 GHz) Dual Core X86 64bit, 2GB DDR3 memory, 8MB SPI Bootflash, 8GB (4GB usable) eMMC bulk flash, and multimode 3G and 4G LTE wireless WAN and IEEE 802.11a/b/g/n WLAN connections. It is resistant to shock, vibration, dust, humidity, and water spray, and a wide temperature range (-40°C to +60°C and type-tested at +85°C for 16 hours) [26].

This mobile edge node comes with two operating systems, a Cisco IOS system that runs a standard Cisco IOS package which handles all routing, switching and networking traffic and a guest operating system IOx running on a virtual machine. The guest operating system IOx runs Linux Yocto which is used to perform the descriptive analytical tasks described in *Section III.D*. However, it is important to point out that developing analytical tasks for the mobile edge nodes involves a trade-off between analytical complexity and processing power. Furthermore, it involves orchestrating highly dynamic, heterogeneous resources at different levels of network hierarchy to support low latency and scalability requirements of transit services.

### C. Data Pre-Processing at the Mobile Edge Node

When the data streams arrive online at the mobile edge node, they are potentially unbounded in size and data tuples may not come in the order. The data streams need to be pre-processed in order to remove errors and inconsistencies. It is very difficult to ensure data quality for the continuous and high volume of data streams, and performing a pre-processing task automatically is even more challenging because the streaming rate is highly dynamic. We implemented a Python script algorithm to handle five automated steps for dealing with *(1) missing tuples*, *(2) duplicated tuples*, *(3) missing attribute values*, *(4) redundant attributes*, and *(5) wrong attribute values*.

*1) Missing tuples:* Too many missing tuples may affect to the final analytical results in the latter stages. So, we eliminated any bus trip that has in total 100 missing tuples and more.

*2) Duplicated tuples:* When the data stream arrived at the edge node, the data tuples is sometimes transmitted twice or more. In this case, any duplicated tuple is automatically found using its timestamp and then removed.

*3) Missing attribute values:* At arriving at the edge, each data tuple contains 17 data fields. However, it may arrive with less than 17. So, we fill up the missing field with "N/A" if it does not involve directly in the latter analytical task. Otherwise, we deleted the whole tuple.

*4) Redundant Attributes:* Opposing with the missing attribute values case, this case happens when new data field is introduced to the data tuple. In this case, the extra attribute is automatically deleted.

*5) Wrong attribute values:* Any attribute might also contain a wrong value due to misspelling, illegal values, and uniqueness violation. In this case, the algorithm first try to standardize the wrong information.

### D. Descriptive Analytics at the Mobile Edge Node

Three descriptive analytical tasks have been developed to reduce the burden on the data hub as well as avoid bottlenecks due to lower network bandwidth. First, each data stream that arrives at the mobile edge node is immediately computed following a sequence of analytical tasks described as one of the following:

*Task 1 - Semantic Annotation:* The aim of this task is to determine whether a bus is moving or not. The GPS coordinates which are sent to the mobile edge node every 5 seconds are used for this computation. In this case, a fixed distance value between two consecutive points is used for determining stops and moves (Fig. 3). This value was empirically determined as being 15m for a transit network. If the distance between the previous point and the current point is more than 15m, the current point is annotated as a move. In contrast, if the distance is less than 15m, the current point is annotated as a stop.

*Task 2 – Temporal Aggregation:* At the end of each trip, this task computes the actual duration and length of the trip, the total number of stops, and the total number of moves. Other data fields such as Trip Identifier, Date, and Start Time are also used for the temporal aggregation. In summary, five data fields *(Trip Id, Date, Start_Time, Total_Move, Total_Stop, Total_Time_Length)* are generated and sent periodically to the data hub located at the Codiac Transit's operation center through the telecommunication network at the end of each trip.

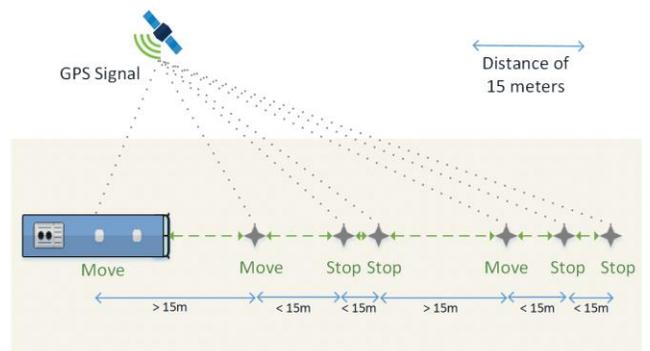

Figure 3. Computation of Moves and Stops of a moving bus

*Task 3 – Summary Function:* It is used to compute the average trip time in the morning (5h-12h), afternoon (13h-18h), and evening (19h-24h). Besides, the average of the total number of moves is also computed for the different times of the day (i.e. morning, afternoon, evening). The same function is also used to compute all the stops. Once the statistics are computed they are sent to the data hub located at the Codiac Transit's operation center through the telecommunication network at the end of the day. During the week, these data could be further analyzed at the data hub for understanding the different mobility patterns during the week.

All the tasks have been implemented in Python 2.7.13. There are several basic libraries exploited such as *time, timedate, csv*. Besides, the library *haversine* was utilized to calculate the great-circle distance between two locations on the Earth surface. The pseudo-code of the algorithm is shown below.

---

**ALGORITHM 1:** Descriptive Analytics at the Mobile Edge Node

---

**Function** *Compute_Move_or_Stop (previous_point, current_point)*
*{*
*if (distance(previous_point, current_point)<15)*
        *return Stop*
*else*
        *return Move*
*}*

*previous_point = null*

**Function** *Process_Data_Stream(incoming_data_point)*
*{*
*current_point = incoming_data_point*
*Compute_Move_or_Stop(previous_point, current_point)*
*previous_point = current_point*

*if (end_of_trip==True)*
*{*
*total_time_length = Compute_Total_Time_Length()*
*total_Move = Compute_Total_Move()*
*total_Stop = Compute_Total_Stop()*
*send_to_the_core( Trip Id, Date, Start_Time, Total_Move,*
                *Total_Stop, Total_Time_Length )*
*}*

*if (end_of_the_day==True)*
*{*
*Average_time_length_morning = total_time_length_morning /*
                *number_of_trips_morning*
*Average_Move_morning = total_Move_morning /*
                *number_of_trips_morning*
*Average_Stop_morning = total_Stop_morning /*
                *number_of_trips_morning*
*Average_time_length_afternoon =total_time_length_afternoon/*
                *number_of_trips_afternoon*
*Average_Move_afternoon = total_Move_afternoon /*
                *number_of_trips_afternoon*
*Average_Stop_afternoon = total_Stop_afternoon /*
                *number_of_trips_afternoon*
*Average_time_length_evening = total_time_length_evening /*
                *number_of_trips_evening*
*Average_Move_evening = total_Move_evening /*
                *number_of_trips_evening*
*Average_Stop_evening = total_Stop_evening /*
                *number_of_trips_evening*
*send_to_the_core ( Average_time_length_morning,*
                *Average_Move_morning,*
                *Average_Stop_morning,*
                *Average_time_length_afternoon,*
                *Average_Move_afternoon,*
                *Average_Stop_afternoon,*
                *Average_time_length_evening,*
                *Average_Move_evening,*
                *Average_Stop_evening )*
*}*
*}*

**Function** *Main()*
*{*
*run_every_5_seconds()*
*Process_Data_Stream(incoming_data_point)*
*}*

---

## IV. RESULTS AND DISCUSSION

In this section, the results from our descriptive analytics platform are visualized to provide actionable information about what is happening in the transit network.

We focus on some prominent abnormalities found in the transit network. Fig. 4 illustrates the existence of several missing trips that have been detected in real-time. Buses did not run on February 14$^{th}$ at 6h to 7h; and there were no trips at 22h on the 15$^{th}$, 16$^{th}$, 18$^{th}$. Moreover, missing trips have also occurred on the 17$^{th}$ after 12h and on the 19$^{th}$ early in the morning (6h and 7h) and in the evening (18h to 22h). This is relevant real-time information for a transit manager to have for an individual bus trip, or a set of trips during a day. For example, the missing trips on February 19$^{th}$ can be explained since it was a Sunday when the Codiac Transit provides a reduced number of trips.

Another interesting result from the descriptive analytics is related to computing a total trip time in real-time. The transit manager can monitor the hourly patterns in real-time and be aware of the outliers. For example, on February 14$^{th}$, the trips have ranged from (897 seconds = 14.95 minutes) to (13,468 seconds = 3.74 hours). But it is also important to point out the occurrence of similar real-time patterns between different days of the week. Some of them might be explained to have occurred due to traffic conditions and snow storms. The transit manager will be able to justify these delays or gather more information to justify such a difference of services being provided.

The transit manager will be also interested in monitoring the total number of moves and stops in real-time and be able to compare them in different times of a day. Fig. 5 shows the variations of the total number of moves and stops belonging to the trips at 8h, 12h, and 16h during the entire one week. The first level of the upper horizontal axis represents the date of a trip, while the second level of the upper horizontal axis shows the three selected departure times at 8h, 12h, and 16h respectively. Additionally, the lower horizontal axis shows the trip identifiers and the vertical axis shows the total number of moves and stops.

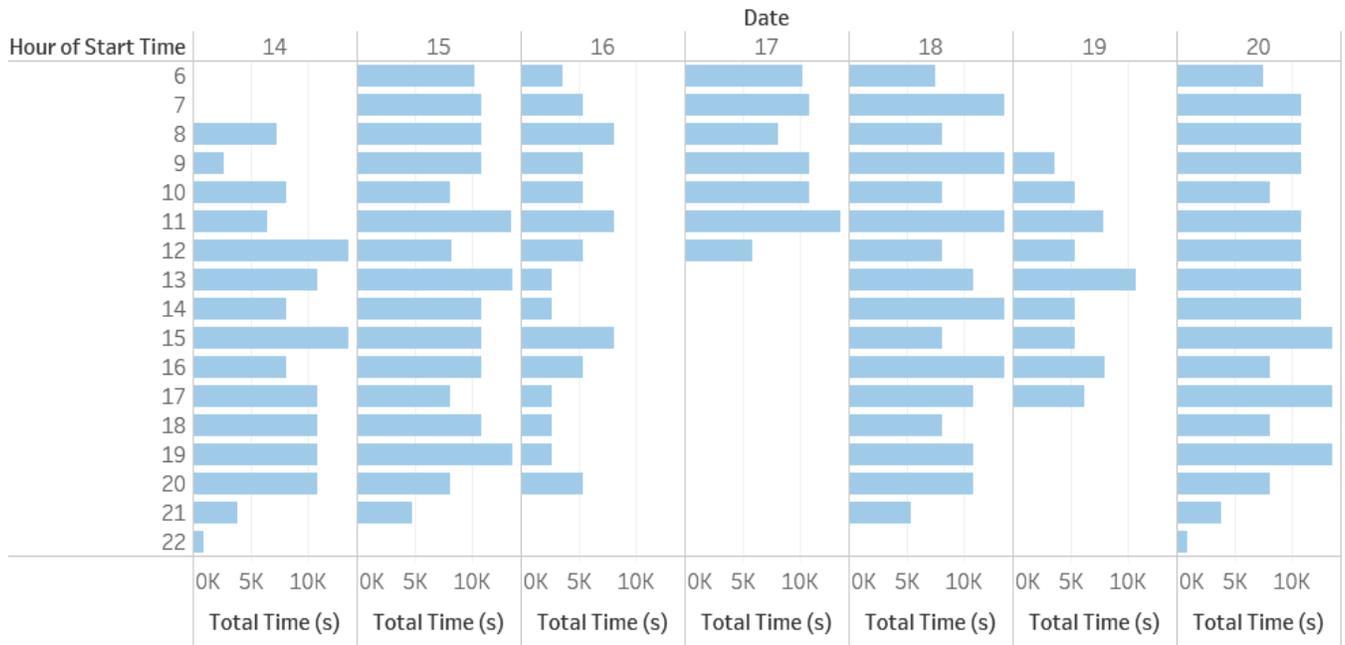

Figure 4. Total trip times during the week

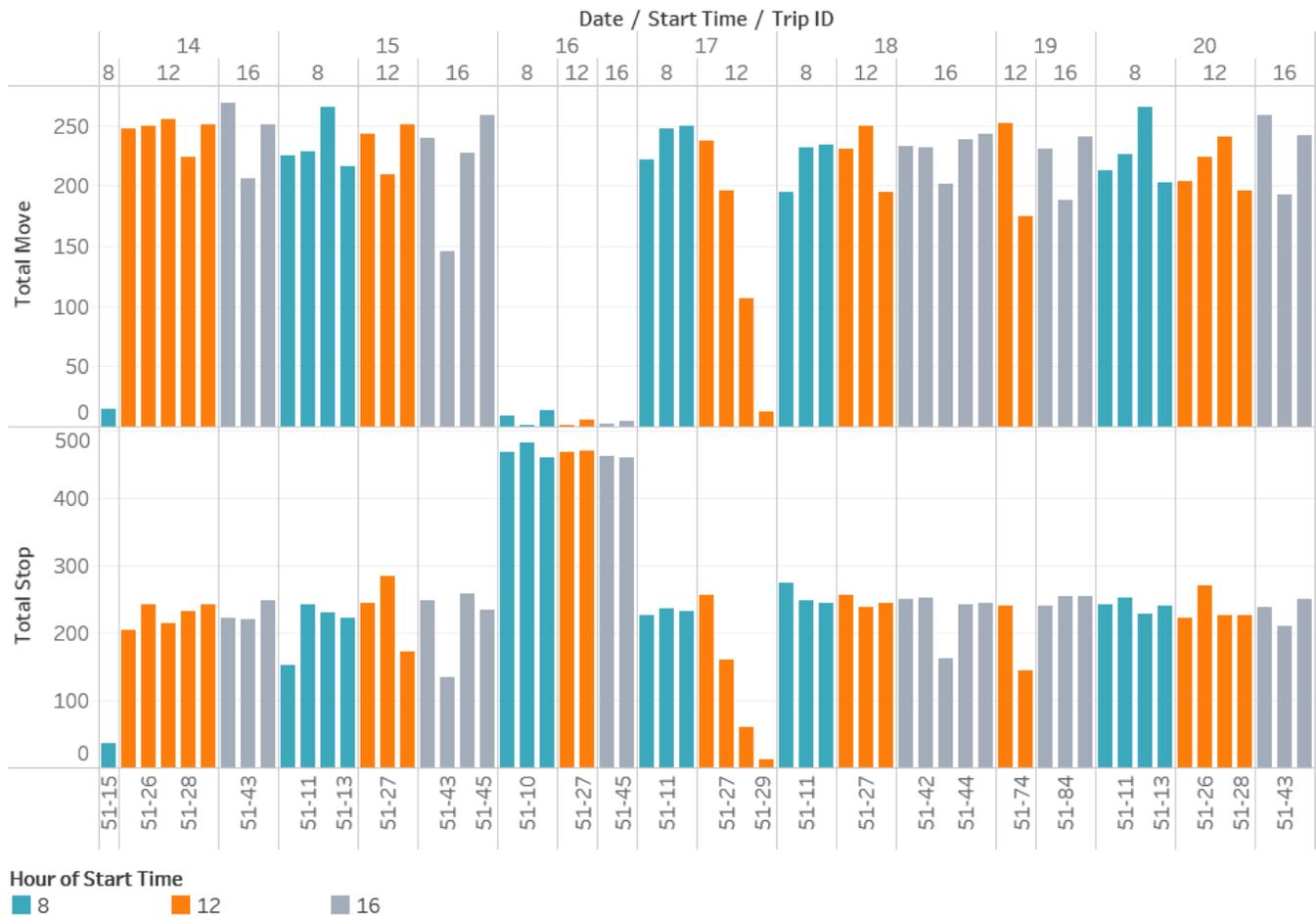

Figure 5. Total number of moves and stops at 8h, 12h, 16h

In general, the distribution of trips is irregular at different times of the day and different days of the week. The patterns show a trend of having more moves than stops during the days, with the exception of February 16th when the number of stops was significantly higher than the moves. This kind of information in real-time can encourage a transit manager to investigate the reason for this situation. In this case, it was a strong snow storm when drivers were stranded in Moncton and major highways were closed. Another aspect that can be useful to transit managers is the fact that the total number of moves is generally similar to the total number of stops except on February 16th. More contextual data is needed for investigating why this pattern is re-occurring.

Fig. 6 shows the total number of trips and the average time of all trips per day. Several outliers can be identified in the graph. In particular, 30 and 31 trips have occurred on February 16th and 17th respectively, in contrast to 47 trips that have occurred on February 14th. The weekdays had usually more than 60 trips (i.e. February 15th, 18th, 20th). We have also computed the average of the total time of all trips per day. Fig. 6 also shows that although the number of trips on February 14th, 16th, 17th, and 19th are lower than other days, the average total time of these days are similar, and some of them even have the highest average total time (i.e. days 14th and 16th). Table 2 provides the results of the descriptive analytics in more detail.

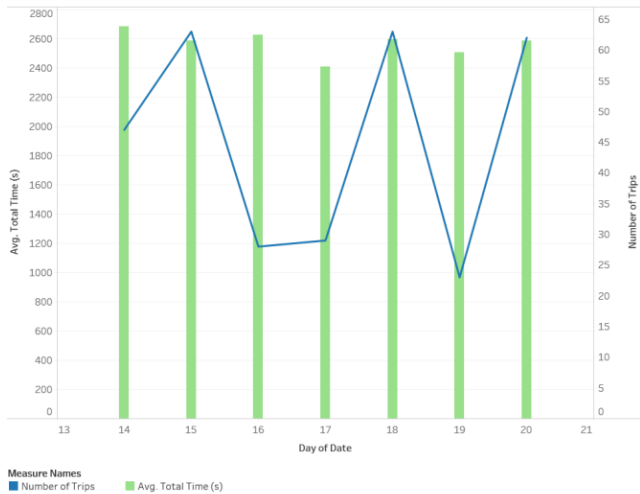

Figure 6. Total number of trips and average trip times

TABLE II. OVERVIEW OF DESCRIPTIVE STATISTICS

| | | Date | | | | | | |
|---|---|---|---|---|---|---|---|---|
| | Period | 14 | 15 | 16 | 17 | 18 | 19 | 20 |
| Average Trip Time (Seconds) | Morning | 3,056 | 2,559 | 2,563 | 2,562 | 2,561 | 2,400 | 2,551 |
| | Evening | 2,393 | 2,390 | 2,691 | | 2,443 | | 2,395 |
| | Afternoon | 2,693 | 2,693 | 2,691 | | 2,693 | 2,532 | 2,692 |
| Average Number of Moves | Morning | 70 | 218 | 10 | 216 | 214 | 191 | 215 |
| | Evening | 224 | 208 | 7 | | 224 | | 203 |
| | Afternoon | 234 | 211 | 2 | | 227 | 214 | 218 |
| Average Number of Stops | Morning | 288 | 225 | 440 | 221 | 234 | 214 | 231 |
| | Evening | 210 | 222 | 473 | | 204 | | 221 |
| | Afternoon | 229 | 233 | 460 | | 230 | 218 | 238 |

Overall, the aggregated statistics are showing that the average number of stops is higher than the average number of moves. Moreover, the average trip time column shows that the average time of a trip is around 40-45 minutes. This is in accordance to the time schedule provided by Codiac Transit. These trends between the moves and stops can be provided in real-time to a transit manager, or in the morning, afternoon, and evening per day (Fig. 7). The patterns reveal how the average number of moves and stops in the afternoons and evenings are very similar, in contrast to the mornings.

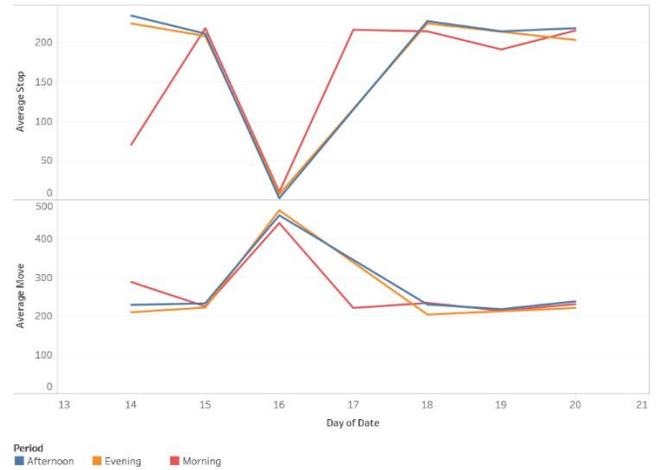

Figure 7. Average number of moves and stops

Finally, we have also produced box and whisker diagrams as illustrated in Fig. 8. Our aim was to detect whether a distribution is skewed and whether there are outliers between the three groups (i.e. morning, afternoon, and evening). The diagrams are showing several outliers that have occurred in the mornings on February 19th and 14th, in the afternoon on February 19th and in the evening on February 16th. Moreover, a transit manger can identify that the drivers usually took more time in the afternoon to finish the trips rather than in the morning and the evening. Additionally, the distribution of average trip time in the afternoon is also constant (around 2,693 seconds = 44.9 minutes). The average trip time in the evening varies from 2,392 seconds (39.8 minutes) to 2,443 seconds (40.7 minutes), in contrast to the median time in the morning of 2,561 seconds (42.7 minutes).

V. CONCLUSIONS

Our experiment has demonstrated the potential of applying edge descriptive analytics for monitoring one bus route. However, the proposed edge computing platform supports the scalability to an entire transit system. It also paves the way to developing new analytical services at the edge network in the near future in order to solve the challenge of fast-growing data produced by the edge devices and sensors.

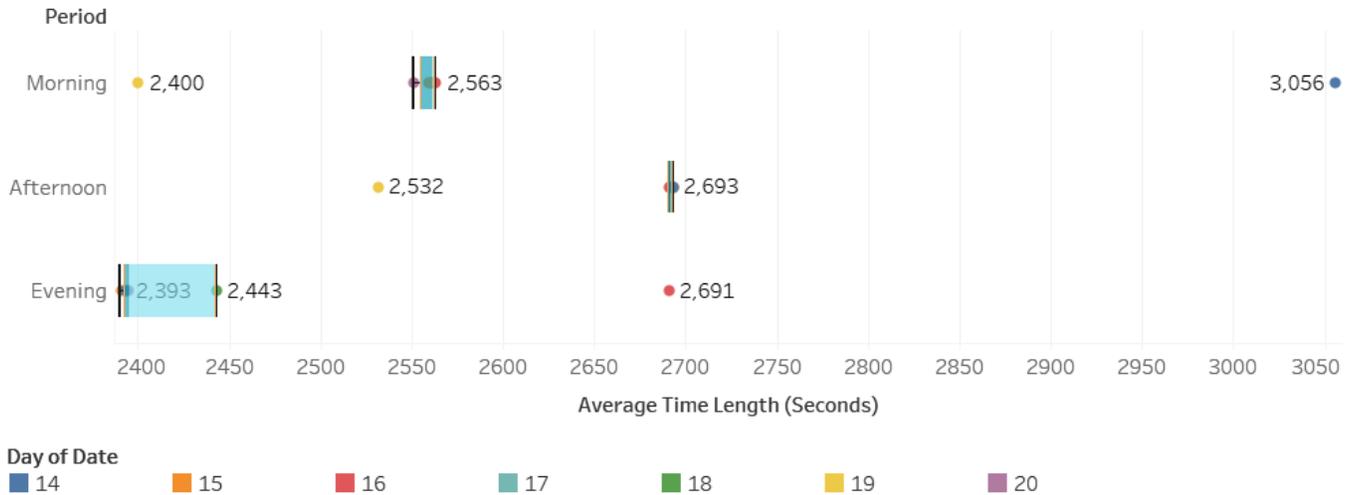

Figure 8. Average trip time

Currently, Codiac Transit does not generate real-time reports of the mobility patterns of their fleet. We have used this experiment to outline the advantages of gaining new insights from real-time descriptive analytics and support Codiac with actionable decision making. But we also see the potential of applying our edge analytical platform in other applications such as autonomous vehicles, smart intersections, and smart traffic light systems.

ACKNOWLEDGMENT

This work was fully supported by the NSERC/Cisco Industrial Research Chair in Real-Time Mobility Analytics. We would like to thank Codiac Transit for providing the data streams. We would also like to thank Nova Communications and Opio Technologies for their support on deploying the mobile edge nodes.